From simes@quark.phy.bnl.gov Mon Mar 22 17:17:05 1999
Date: Mon, 22 Mar 1999 17:12:36 -0500 (EST)
From: "Fern M. Simes" <simes@quark.phy.bnl.gov>

\documentstyle[prd,aps,preprint]{revtex}

\newcommand{\ssthwmzms}{\sin^2\theta_W(m_Z)_{\overline{MS}}}
\newcommand{\ssthwz}{\sin^2\theta_W^0}
\newcommand{\mzms}{(m_Z)_{\overline{MS}}}
\newcommand{\beq}{\begin{equation}}
\newcommand{\eeq}{\end{equation}}
\newcommand{\beqa}{\begin{eqnarray}}
\newcommand{\eeqa}{\end{eqnarray}}
\newcommand{\bml}{\begin{mathletters}}
\newcommand{\eml}{\end{mathletters}}
\newcommand{\noi}{\noindent}

\newcommand{\lsim}{\mathrel{\lower4pt\hbox{$\sim$}}
\hskip-11pt\raise1.6pt\hbox{$<$}\;}

\newcommand{\gsim}{\mathrel{\lower4pt\hbox{$\sim$}}
\hskip-12.5pt\raise1.6pt\hbox{$>$}\;}

\newcommand{\lsims}{\mathrel{\lower4pt\hbox{$\sim$}}
\hskip-10.5pt\raise1.6pt\hbox{$<$}\;}

\newcommand{\gsims}{\mathrel{\lower4pt\hbox{$\sim$}}
\hskip-10.5pt\raise1.6pt\hbox{$>$}\;}
\begin{document}

\title{Fermi Constants and ``New Physics''}

\author{William J. Marciano}
\address{Brookhaven National Laboratory \\
Upton, New York\ \ 11973}
\maketitle

\begin{abstract}
Various precision determinations of the Fermi constant are compared.
Included are muon and (leptonic) tau decays as well as indirect
prescriptions employing $\alpha$, $m_Z$, $m_W$, $\ssthwmzms$,
$\Gamma(Z\to\ell^+\ell^-)$, and $\Gamma(Z\to \nu\bar \nu)$ as input.
Their good agreement tests the standard model at the $\pm0.1\%$ level
and provides stringent constraints on new physics. That utility is
illustrated for: heavy neutrino mixing, 2 Higgs doublet models, $S$,
$T$, and $U$ parameters and excited $W^{\ast^\pm}$ bosons (Kaluza-Klein
excitations). For the last of those examples, $m_{W^\ast}\gsims 2.9$
TeV is found.
\end{abstract}

\vspace{22pt}

The Fermi constant, $G_F$, is an important, venerable holdover from the
old local theory of weak interactions \cite{zp88161}. Expressed in
terms of SU(2)$_L\times{}$U(1)$_Y$ standard model parameters, it is
 given by

\beq
G_F = g^2_2/4\sqrt{2} m^2_W \label{eq1}
\eeq

\noi where $g_2$ is an SU(2)$_L$ gauge coupling and $m_W$ is the
$W^\pm$ gauge boson mass. To be more precise, $G_F$ must be expressed
in terms of physical observables or well prescribed renormalized
parameters. Also, electroweak radiative corrections must be properly
accounted for.

Traditionally, the muon lifetime, $\tau_\mu$, has been used to define
the Fermi constant because of its very precise experimental value
\cite{epjc31} 

\beq
\tau_\mu = 2.197035 (40) \times 10^{-6} s \label{eq2}
\eeq

\noi and theoretical simplicity. Labeling that definition by $G_\mu$,
it is related to $\tau_\mu$ via \cite{ap2026}

\beqa
\tau^{-1}_\mu & = & \Gamma (\mu\to {\rm all}) = \frac{G^2_\mu
m^5_\mu}{192\pi^3} f \left(\frac{m^2_e}{m^2_\mu} \right) (1+R.C.)
\left( 1+\frac{3}{5}\, \frac{m^2_\mu}{m^2_W} \right) \nonumber \\
f(x) & = & 1-8x +8x^3 -x^4 -12x^2 \ell nx \label{eq3}
\eeqa

\noi In that expression, R.C. stands for radiative corrections and the
$\frac{3}{5} m^2_\mu/m^2_W$ term is a small  $W$ boson
propagator effect. The R.C. expression is somewhat arbitrary. Most
quantum loop corrections to muon decay are absorbed into the
renormalized parameter $G_\mu$. For historical reasons and in the
spirit of effective field theories, R.C. is defined to be the QED
radiative corrections to muon decay in the local V-A four fermion
description of muon decay. That separation is natural and practical,
since those QED corrections are finite to all orders in perturbation
theory \cite{ap2026}. In fact, they have been fully computed through ${\cal
O}(\alpha^2)$ and are given by

\beq
R.C. = \frac{\alpha}{2\pi} \left(\frac{25}{4} -\pi^2\right) \left( 1
+\frac{\alpha}{\pi} \left( \frac{2}{3} \ell n\frac{m_\mu}{m_e} -3.7
\right) + \left( \frac{\alpha}{\pi}\right)^2 \left( \frac{4}{9} \ell n^2
\frac{m_\mu}{m_e} -2.0 \ell n \frac{m_\mu}{m_e} + C\right) +\cdots
\right) \label{eq4}
\eeq

\noi where $\alpha$ is the fine structure constant

\beq
\alpha^{-1} = 137.03599959 (40) \label{eq5}
\eeq

\noi The leading ${\cal O}(\alpha)$ term in that expression has been
known for 4 decades from the pioneering work of Kinoshita and Sirlin
\cite{pr1131652} and Berman \cite{pr112267}. Coefficients of higher
order $\ell n \frac{m_\mu}{m_e}$ terms are determined by the
renormalization group requirement \cite{npb29296}

\beqa
& \left( m_e\frac{\partial}{\partial m_e} + \beta (\alpha)
\frac{\partial}{\partial \alpha} \right) R.C. = 0 & \nonumber \\
& \beta(\alpha) = \frac{2}{3}\, \frac{\alpha^2}{\pi} + \frac{1}{2}\,
\frac{\alpha^3}{\pi^2} + \cdots& \label{eq6}
\eeqa

\noi The -3.7 two loop term was very recently computed by van Ritbergen
and Stuart \cite{prl82488}. Their result also implies the
next-to-leading logs in (\ref{eq4}) via (\ref{eq6}), leaving $C$ as the
only unknown ${\cal O}(\alpha^3)$ contribution to R.C\null. Comparing
(\ref{eq3}) and (\ref{eq2}), one finds

\beq 
G_\mu = 1.16637(1) \times 10^{-5} {\rm GeV}^{-2} \label{eq7}
\eeq

\noi which is, by far, the best determination of the Fermi constant. In
fact, it is more than 100 times better than the other prescriptions
considered in this paper. Nevertheless, there have been several
proposals to further reduce the uncertainty in $\tau_\mu$ and $G_\mu$
by an additional factor of 10. Given the fundamental nature of $G_\mu$,
such measurements should certainly be encouraged. However, from the
point of view of testing the standard model, some other independent
determination of the Fermi constant would have to catch up to $G_\mu$
before a more precise $\tau_\mu$ measurement could be fully utilized.

In the renormalization of $G_\mu$, lots of interesting quantum loop
effects have been absorbed. Included are top and Higgs loop corrections
to the $W$ boson propagator as well as potential new physics from
supersymmetry, technicolor, etc. Even possible tree level
contributions, for example from massive excited $W^{\ast^\pm}$ bosons
or other effects, might be encoded in $G_\mu$. To unveil such
contributions requires comparison of $G_\mu$ with other independent
determinations of the Fermi constant that could have different tree or loop
level dependences.

Because of the renormalizability of the standard model, universality of
bare gauge couplings among lepton generations \cite{prd83612}

\beq
g^e_{2_0} = g^\mu_{2_0} = g^\tau_{2_0} \label{eq8}
\eeq

\noi and the bare natural relations \cite{nc16a423}

\beq
\ssthwz = \frac{e^2_0}{g^2_{2_0}} = 1- (m^0_W/m^0_Z)^2, \label{eq9}
\eeq

\noi there are many ways to determine Fermi constants and compute very
precisely their relationships with $G_\mu$. Comparison of those
quantities can then be used to test the standard model and probe for
new physics.

The leptonic decay widths of the tau can provide, in close analogy with
muon decay, Fermi constants $G_{\tau\ell}, \ell = e$ or $\mu$.
Including ${\cal O}(\alpha)$ QED corrections, one employs the radiative
inclusive rate \cite{prl611815}

\beq
\Gamma(\tau\to\ell \nu\bar \nu(\gamma)) =
\frac{G^2_{\tau\ell}m^5_\tau}{192\pi^3} f
\left(\frac{m^2_\ell}{m^2_\tau}\right) \left( 1+ \frac{3}{5}\,
\frac{m^2_\tau}{m^2_W} \right) \left( 1+ \frac{\alpha}{2\pi} \left(
\frac{25}{4} - \pi^2\right)\right) \label{eq10}
\eeq

\noi Those Fermi constants have been normalized, through ${\cal
O}(\alpha)$, such that $G_{\tau e} = G_{\tau\mu} = G_\mu$ in the
standard model. That is possible because $g^e_{2_0} = g^\mu_{2_0} =
g^\tau_{2_0}$ and the ${\cal O}(\alpha)$ radiative corrections are the
same (up to ${\cal O} (\alpha m^2_\tau/m^2_W))$. 

Employing the experimental averages \cite{hepph9812399} 

\bml
\label{eq11}
\beq
\tau_\tau = 290.5\pm 1.0\times10^{-15} s \label{eq11a}
\eeq
\beq
B(\tau\to e\nu\bar\nu(\gamma)) = 0.1781(6) \label{eq11b}
\eeq
\beq
B(\tau\to\mu\nu\bar\nu(\gamma)) = 0.1736(6) \label{eq11c}
\eeq
\eml

\noi implies

\bml
\label{eq12}
\beqa
\Gamma (\tau \to e \nu\bar\nu(\gamma)) & = & 4.035 (19)\times10^{-13}
{\rm ~GeV} \label{eq12a} \\
\Gamma(\tau\to\mu\nu\bar\nu (\gamma)) & = & 3.933 (19) \times 10^{-13}
{\rm ~GeV} \label{eq12b}
\eeqa
\eml

\noi Used in conjunction with

\beq 
m_\tau = 1777.0 (3) {\rm ~MeV}, \label{eq13}
\eeq

\noi those widths lead to 

\beqa
G_{\tau e} & = & 1.1666 (28) \times 10^{-5} {\rm ~GeV}^{-2}
\label{eq14} \\
G_{\tau\mu} & = & 1.1679 (28) \times 10^{-5} {\rm ~GeV}^{-2}
\label{eq15}
\eeqa

\noi They are in very good accord with $G_\mu$, but their errors are
nearly 300 times larger. Nevertheless, collectively those Fermi
constants test $e$-$\mu$-$\tau$ universality at the $\pm0.2\%$ level

\beq
g^e_{2_0} : g^\mu_{2_0} : g^\tau_{2_0} :: 1 : 1.0011 (24) : 1.0006 (24)
\label{eq16}
\eeq

\noi (employing (\ref{eq11b}) and (\ref{eq11c}) directly).

The good agreement between $G_\mu$ and the $G_{\tau\ell}$ can be used
to constrain new physics. Consider, for example, the effect of a heavy
fourth generation lepton doublet $(\nu_4, L)$ with masses $\gsim 95$
GeV; so, it would have escaped detection at existing colliders.
Parametrizing the 3rd and 4th generation mixing by $\theta_{34}$, one
has (assuming no mixing with the first or second generations)
\cite{npb403,pr1571,pr45r721} 

\beq
\nu_\tau = \nu_3\cos\theta_{34} + \nu_4\sin\theta_{34} \label{eq17}
\eeq

\noi That being the only mixing effect, one would expect $G_{\tau\ell}
= G_\mu\cos\theta_{34}$. Combining (\ref{eq14}) and (\ref{eq15}) to get
$G^{\rm ave}_{\tau\ell} = 1.1672 (25) \times 10^{-5}$ GeV$^{-2}$ and
comparing with $G_\mu$, one finds the rather stringent bound

\beq
\sin\theta_{34} <0.075 \qquad (95\% {\rm CL}) \label{eq18}
\eeq

\noi What value of $\sin\theta_{34}$ might be reasonable in such a
scenario? If an analogy with quark mixing is appropriate, one might
guess \cite{npb403} $\sin\theta_{34} \approx \sqrt{m_\tau/m_L}$. If
that is the case, (\ref{eq18}) translates to $m_L\gsim 316$ GeV\null.
An additional factor of 2 improvement in $G_{\tau\ell}$ would push that
probe into the very interesting $m_L\gsim 850$ GeV region, under the
above assumptions. A similar analysis could be applied to singlet
neutrinos or more general mixing scenarios. Note, however, that heavy
$\nu_4$ mixing with the first two generations of neutrinos must be
suppressed due to constraints from $\mu\to e\gamma$ and $\mu^- N\to e^-
N$ searches.

As a second illustration of new physics, consider the general 2 Higgs
doublet model with $\tan \beta = v_2/v_1$ and physical scalar masses
$m_h$, $m_H$, $m_A$ and $m_{H^\pm}$. Charged Higgs scalar exchange at
the tree level would reduce the tau leptonic decay rates by a factor
\cite{plb333403} $\left(1-\frac{2m^2_\ell}{m^2_{H^\pm}}
\tan^2\beta\right)$ and thus effectively imply $G_{\tau\mu} <G_{\tau
e}$. However, the good agreement between (\ref{eq15}) and (\ref{eq14})
can be used to set the bound \cite{npb403}

\beq
m_{H^\pm} \gsim 2\tan \beta {\rm ~GeV} \qquad (95\% {\rm ~CL})
\label{eq19}
\eeq

\noi For large $\tan \beta\gsim 45$, that bound is competitive with
direct $e^+e^-$ collider searches as well as constraints from
$B\to\tau\nu X$ \cite{prd482342}. However, $b\to s\gamma$ measurements
generally give a more restrictive bound. Constraints on the spectrum of
scalars can also be obtained by comparing $G_\mu$ and $G_{\tau\ell}$,
but they will not be discussed here \cite{plb28575}.

There are also a number of indirect prescriptions for obtaining Fermi
constants. For example, using the natural relations in (\ref{eq9}), one
can define \cite{npb84132,prd22471}

\beqa
G^{(1)}_F & = & \frac{\pi\alpha}{\sqrt{2} m^2_W (1-m^2_W/m^2_Z)
(1-\Delta r)} \label{eq20} \\
G^{(2)}_F & = & \frac{\pi\alpha}{\sqrt{2} m^2_W\ssthwmzms (1-\Delta r
\mzms)} \label{eq21} \\
G^{(3)}_F & = & \frac{4\pi\alpha}{\sqrt{2} m^2_Z\sin^22\theta_W\mzms
(1-\Delta\hat r)} \label{eq22}
\eeqa

\noi where $\Delta r$, $\Delta r\mzms$, and $\Delta\hat r$ represent
the radiative corrections to those relationships. They have been
normalized such that $G_\mu= G^{(1)}_F = G^{(2)}_F = G^{(3)}_F$ in the
standard model \cite{rep444p35}. To determine those quantities,
requires calculations of the loop corrections to $G_\mu$, $\alpha$,
$m_Z$, $m_W$, and $\ssthwmzms$ as well as the reactions used to measure
them. Fortunately, the complete one loop corrections in
(\ref{eq20})--(\ref{eq22}) are known and most leading higher order
effects have also been computed \cite{plb394188}.

Leptonic partial widths of the $Z$ boson also provide useful Fermi
constant determinations. Defining

\beqa
G^{Z\ell^+\ell^-}_F & = & \frac{12\sqrt{2} \pi\Gamma (Z\to \ell^+\ell^-
(\gamma))}{m^3_Z (1-4\ssthwmzms + 8 \sin^4\theta_W\mzms) (1-\Delta
r_Z\mzms)} \label{eq23} \\
G^{Z\nu\bar\nu}_F & = & \frac{4\sqrt{2} \pi\Gamma (Z\to
\Sigma\nu\bar\nu)}{m^3_Z (1-\Delta r_Z)} \label{eq24}
\eeqa

\noi with the radiative corrections $\Delta r_Z\mzms$ and $\Delta r_Z$
again normalized such that $G^{Z\ell^+\ell^-}_F = G^{Z\nu\bar\nu}_F =
G_\mu$ in the standard model. Note that $\Gamma(Z\to
\ell^+\ell^-(\gamma))$ by definition  corresponds to $Z$ decay into
massless charged 
leptons \cite{cern97154}, $m_\ell=0$, It is obtained from an average of
$\ell=e$, $\mu$, $\tau$ data, where only the $\tau^+\tau^-$ width
requires a non-negligible phase space correction factor of 1.0023. For
some new physics scenarios \cite{zpc51695}, a separate
$G^{Z\tau^+\tau^-}_F$ could prove useful; however, those cases will not
be considered here.

The electroweak radiative corrections in (\ref{eq20})--(\ref{eq24}) are
known. They depend with varying sensitivities on the top quark and
Higgs masses. For example, $\Delta r\mzms$ exhibits very little
dependence on those quantities while $\Delta r$ is most sensitive. Also,
the first three, $\Delta r$, $\Delta r\mzms$, and $\Delta\hat r$ have a
common low energy hadronic vacuum polarization loop uncertainty \cite{prd20274} due to
$\alpha$. Here, a very small $\pm0.0002$ error from that source is
assigned \cite{plb439427}. A more conservative approach might expand
\cite{zpc67585} that uncertainty by a factor of 2--4, but it would not
affect our subsequent analysis significantly.

In the evaluation of electroweak radiative corrections, the following
central values and uncertainty ranges are assumed

\beqa
m_t & = & 174.3\pm5.1 {\rm ~GeV} \nonumber \\
m_H & = & 125^{+275}_{-35} {\rm ~GeV} \label{eq25}
\eeqa

\noi The Higgs mass range is bounded from below by LEP II results
$m_H\gsim 89.8$ GeV\null. A conservative upper range of $m_H\sim400$
GeV is assumed at the 1 sigma level. Using those input parameters, one
finds \cite{plb394188}

\bml
\label{eq26}
\beqa
\Delta r & = & 0.0358\mp 0.0020^{+0.0049}_{-0.0012} \pm 0.0002
\label{eq26a} \\
\Delta r\mzms & = & 0.0696\pm 0.0001^{+0.0005}_{-0.0003} \pm 0.0002
\label{eq26b} \\
\Delta\hat r & = & 0.0597\mp 0.0005^{+0.0017}_{-0.0005} \pm 0.0002
\label{eq26c} \\
\Delta r_Z\mzms & = & -0.0071\mp 0.0005^{+0.0008}_{-0.0001}
\label{eq26d} \\
\Delta r_Z & = & -0.0048\mp 0.0005^{+0.0008}_{-0.0001} \label{eq26e}
\eeqa
\eml

\noi where the first error corresponds to $\Delta m_t$, the second to
$\Delta m_H$, and the third (when present) to hadronic vacuum
polarization uncertainties. Increasing the last of those by a factor of
2--4 would make it comparable to other errors in $\Delta r\mzms$ and
$\Delta\hat r$, but would not seriously impact our subsequent results.

Employing the values of $\alpha$, $m_t$, and $m_H$ given above, along
with \cite{slactop}

\bml
\label{eq27}
\beqa
m_Z & = & 91.1867 (21) {\rm ~GeV} \label{27a} \\
m_W & = & 80.422 (49) {\rm ~GeV} \label{27b} \\
\ssthwmzms & = & \sin^2\theta^{\rm eff}_W -0.00028 = 0.23100 (22)
\label{eq27c} \\
\Gamma (Z\to \ell^+\ell^-(\gamma)) & = & 83.91 (10) {\rm ~MeV}
\label{eq27d} \\
\Gamma (Z\to\Sigma\nu\bar\nu) & = & 500.1 (18) {\rm ~MeV} \label{eq27e}
\eeqa
\eml

\noi leads to 

\bml
\label{eq28}
\beqa
G^{(1)}_F & = & 1.1700 (\mp 0.0036)(^{+0.0062}_{-0.0027})\times 10^{-5}
{\rm ~GeV}^{-2} \label{eq28a} \\
G^{(2)}_F & = & 1.1661 (\mp 0.0018) (^{+0.0005}_{-0.0004}) \times 10^{-5}
{\rm ~GeV}^{-2} \label{eq28b} \\
G^{(3)}_F & = & 1.1672 (\mp 0.0008) (^{+0.0018}_{-0.0007}) \times 10^{-5}
{\rm ~GeV}^{-2} \label{eq28c} \\
G^{Z\ell^+\ell^-}_F & = & 1.1650 (\pm0.0014) (^{+0.0011}_{-0.0006})
\times 10^{-5} {\rm ~GeV}^{-2} \label{eq28d} \\
G^{Z\nu\bar\nu}_F & = & 1.1666 (\pm0.0042)(^{+0.0011}_{-0.0006}) \times 10^{-5}
{\rm ~GeV}^{-2} \label{eq28e} 
\eeqa
\eml

\noi where the first error comes from the experimental input in
(\ref{eq27}) while the second is due to uncertainties in (\ref{eq26})
from radiative corrections.

All derived Fermi constants in (\ref{eq28}) are in excellent accord
with $G_\mu = 1.16637 (1) \times 10^{-5}$ GeV$^{-2}$, but their errors
are more than 100 times larger. Nevertheless, they can be used to place
tight constraints on new physics.

Consider the case of heavy new chiral SU(2)$_L$ doublets from a fourth
generation of fermions or motivated by technicolor models of dynamical
electroweak symmetry breaking. Such fermions contribute to the above
radiative corrections via gauge boson self-energies. Their effects are
conveniently described by the Peskin-Takeuchi $S$, $T$, and $U$
parameters \cite{prl65964}, where $S$ represents isospin-conserving and
$T$ and $U$ isospin-violating gauge boson loop contributions. Their
presence would 
modify the relationships between $G_\mu$ and the other Fermi constants
such that

\bml
\label{eq29}
\beqa
G_\mu & = & G^{(1)}_F (1+0.017 S-0.026 T-0.020 U) \label{eq29a} \\
G_\mu & = & G^{(2)}_F (1+0.0085 (S + U)) \label{eq29b} \\
G_\mu & = & G^{(3)}_F (1+0.011 S-0.0078 T) \label{eq29c} \\
G_\mu & = & G^{Z\ell^+\ell^-}_F (1-0.0078 T) \label{eq29d} \\
G_\mu & = & G^{Z\nu\bar\nu}_F (1-0.0078T) \label{eq29e}
\eeqa
\eml

\noi No evidence for $S$, $T$, or $U\ne0$ is apparent from
(\ref{eq28}). In fact, comparing (\ref{eq29b}) with $G_\mu$ and
$G^{(2)}_F$ in (\ref{eq28b}) leads to 

\beq
-0.28 < S+U<0.33 \qquad (90\% {\rm~CL}) \label{eq30}
\eeq

\noi Comparing (\ref{eq29c}) with (\ref{eq29d}) and (\ref{eq29e})
eliminates the dependence on $T$ and gives the somewhat tighter
constraint

\beq
-0.38 < S < 0.04 \qquad (90\% {\rm ~CL}) \label{eq31}
\eeq

In the case of a heavy fourth generation of fermions (4 chiral
doublets), one expects 
$S=2/3\pi\simeq 0.21$ which conflicts with (\ref{eq31}). Generic
technicolor models suggest \cite{prl65964} $S\sim {\cal O}(+1)$ which
conflicts significantly with (\ref{eq31}) and (\ref{eq30}) for $U\simeq
0$ (as expected in those models). The bound on $S$ provides an obstacle
for electroweak dynamical symmetry breaking advocates or fourth
generation scenarios. If high mass chiral fermion doublets exist, their
dynamics must exhibit properties that preserve $S\sim0$ or other loop
effects must cause a cancellation.

>From the comparison of (\ref{eq29d}) and (\ref{eq29e}) with $G_\mu$,
one also obtains the bound

\beq
-0.40 < T < 0.17 \qquad (90\% {\rm ~CL}) \label{eq32}
\eeq

\noi on the isospin violating loop correction. The constraints in
(\ref{eq30})--(\ref{eq32}) are nearly as good as those obtained from
global fits to all electroweak data \cite{hepph9809352}.

The final example considered here is the possibility of excited
$W^{\ast^\pm}$ bosons that arise in theories with extra compact
dimensions (Kaluza-Klein excitations) \cite{plb27021} or models with
composite gauge bosons. Assuming fermionic couplings to $W^{\ast^\pm}$
identical to those of the $W^\pm$, $g^\ast_2 = g_2 $, direct searches
at the Tevatron lead to the bound \cite{epjc31}

\beq
m_{W^\ast} > 720 {\rm ~GeV} \qquad (95\% {\rm ~CL}) \label{eq33}
\eeq

\noi For $g^\ast_2\ne g_2$, that bound is (roughly) multiplied by
$1+0.3 \ell n (g^\ast_2/g_2)$ and thus not so sensitive to shifts in
$g^\ast_2$. If such bosons exist, they would also 
contribute to low energy charged current amplitudes such as muon or tau
decays. Their effect would be encoded in $G_\mu$ and $G_{\tau\ell}$
but not the indirect Fermi constants in (\ref{eq28}).

The effect of excited bosons would be to replace $g^2_2/m^2_W$ in low
energy amplitudes by $g^2_2/\langle m^2_W\rangle$ where \cite{slactop}

\beq
\frac{1}{\langle m^2_W\rangle} = \frac{1}{m^2_W}
+\frac{(g^\ast_2/g_2)^2}{m^2_{W^\ast}} + \frac{(g^{\ast\ast}_2
/g_2)^2}{m^2_{W^{\ast\ast}}} + \cdots \label{eq34}
\eeq

\noi As long as the relative signs are positive, $\langle m^2_W\rangle$
is always smaller than $m^2_W$. The situation is analogous to adding
resistors in parallel. In such a scenario, $G_\mu$ should be larger
than the $G_F$ in (\ref{eq28}). There is no indication of such an
effect. Quantitatively, one expects

\beqa
G_\mu & = & G^{(i)}_F \left( 1+ C \left(\frac{g^\ast_2}{g_2}\right)^2
\frac{m^2_W}{m^2_{W^\ast}} \right) \nonumber \\
C & = & 1 + \left(\frac{g^{\ast\ast}_2}{g^\ast_2} \right)^2
\frac{m^2_{W^\ast}}{m^2_{W^{\ast\ast}}} + \cdots >1 \label{eq35}
\eeqa

\noi In the simplest single extra dimension theory \cite{plb27021},
$C\simeq \sum\limits^\infty_{n=1} 1/n^2 = \pi^2/6$. Additional compact space
dimensions can further increase $C$.

Comparing (\ref{eq35}) with (\ref{eq28}), one finds (at 95\% CL)

\bml
\label{eq36}
\beqa
m_{W^\ast} & > & 2.9 \sqrt{C} (g^\ast_2/g_2) {\rm ~TeV} \qquad ({\rm
from~} G^{(3)}_F) \label{eq36a} \\
m_{W^\ast} & > & 1.5 \sqrt{C} (g^\ast_2/g_2) {\rm ~TeV} \qquad ({\rm
from~} G^{(2)}_F) \label{eq36b} \\
m_{W^\ast} & > & 1.4 \sqrt{C} (g^\ast_2/g_2) {\rm ~TeV} \qquad ({\rm
from~} G^{(1)}_F) \label{eq36c} \\
m_{W^\ast} & > & 1.4 \sqrt{C} (g^\ast_2/g_2) {\rm ~TeV} \qquad ({\rm
from~} G^{Z\ell^+\ell^-}_F) \label{eq36d} \\
m_{W^\ast} & > & 1.0 \sqrt{C} (g^\ast_2/g_2) {\rm ~TeV} \qquad ({\rm
from~} G^{Z\nu\bar\nu}_F) \label{eq36e} 
\eeqa
\eml

\noi Note that $G^{Z\ell^+\ell^-}_F$ would lead to a better bound if
its central value were not about 1 sigma below $G_\mu$. Also, the bound
from $G^{(2)}_F$ has less $m_t$ and $m_H$ sensitivity and  probably
provides the least model dependent constraint.

The above bounds can be relaxed if $g^\ast_2\ll g_2$ or increased for
$C>1$. Taking $m_{W^\ast}>2.9$ TeV as representative, that corresponds
to a bound on $W^\pm$ substructure at $\sim2.9$ TeV and $R<
1/m_{W^\ast} \sim 7\times 10^{-18}$ cm for the radii of extra
dimensions \cite{hepph9902323}.

How might the above constraints improve? Measurement of $m_H$ and
refinements in $m_t$ will reduce the uncertainty in radiative
corrections. At LEP II and the Tevatron, a reduction in $\Delta m_W$ to
$\pm15$ MeV is anticipated while at SLC, $\Delta \ssthwmzms$ could be
reduced to $\pm0.00018$. In the longer term, high statistics $Z$ pole
studies at a future $\ell^+\ell^-$ collider could reduce
$\Delta\ssthwmzms$ to about $\pm0.00004$ and significantly improve the
leptonic $Z$ partial widths. Such improvements will, for example, allow
one to probe $m_{W^\ast}$ beyond $5 \sqrt{C}(g^\ast_2/g_2)$ TeV\null. For
comparison, direct searches at the Tevatron with $2fb^{-1}$ will
explore $m_{W^\ast}\lsim1.2$ TeV while LHC is sensitive to $\sim6$
TeV\null. An advantage of direct collider searches for excited bosons
is their reduced sensitivity to changes in $g^\ast_2$, as long as their
leptonic branching ratio remains relatively fixed and is significant. On
the other hand, indirect constraints obtained by comparing $G_\mu$ and
$G^{(i)}_F$ are more sensitive to $g^\ast_2$, but independent of
branching ratio assumptions. Hence, the two approaches are very
complementary. 

In addition to the above, one can define Fermi constants using quark
beta decays and CKM unitarity or from low energy neutral current
processes such as atomic parity violation. The latter case provides a
powerful constraint on many examples of new physics. It will be
examined in a subsequent paper which updates the radiative corrections
to atomic parity violation.

The Fermi constant has played an important role in the history of weak
interactions and development of the standard model. As demonstrated
here, it continues to provide useful guidance for testing the
standard model and probing new physics.

\end{document}